\documentclass[doublecol]{epl2} 

\usepackage{amsmath,amssymb,amsfonts}
\usepackage{graphics,graphicx}

\newcommand{\Zeff}{Z_{\text{up}}}
\newcommand{\dmax}{\delta_{\text{max}}}
\newcommand{\Qint}{Q_{\text{int}}}

\newcommand{\be}{\begin{equation}}
\newcommand{\ee}{\end{equation}}

\begin{document}

\title{Strong screening in the plum pudding model} 

\author{A.D. Chepelianskii\inst{1}, F. Closa\inst{2}, E. Rapha\"el\inst{2} 
and E. Trizac\inst{3}} 
\shortauthor{Chepelianskii \etal}

\institute{                    
  \inst{1} LPS, UMR CNRS 8502, B\^at. 510, Universit\'e 
Paris-Sud, 91405 Orsay, France\\
\inst{2} Gulliver, 
UMR CNRS 7083, ESPCI, 10 rue Vauquelin, 75005 Paris, France   \\
   \inst{3} Universit\'e Paris-Sud, Laboratoire de
Physique Th\'eorique et Mod\`eles Statistiques, UMR CNRS 8626, 91405 Orsay, 
France
}

\pacs{82.70.Dd}{}
\pacs{52.27.Gr}{}
\pacs{05.20.Jj}{}

\abstract{
We study a generalized Thomson problem that appears in several condensed matter
settings: identical point-charge particles can penetrate inside a 
homogeneously charged sphere, with global electro-neutrality.
The emphasis is on scaling laws at large Coulombic couplings, and deviations
from mean-field behaviour, by a combination of Monte Carlo simulations and
an analytical treatment within a quasi-localized charge approximation, 
which provides reliable predictions. We also uncover a local overcharging
phenomenon driven by ionic correlations alone.}


\date{\today}
\maketitle

The venerable Thomson problem of finding the ground state of an ensemble
of electrons confined in an homogeneously charged neutralizing sphere, 
is still unsolved and has a long history, see e.g.  
\cite{T1904,Thomson_suite,Thomson_variants} and references therein
for the different generalizations that have been put forward.
The model was introduced
at the beginning of the 20th century \cite{T1904},
just after the discovery of the electron, 
but before that of the proton,  as a classical representation of the atom;
hence the ''plums'' representing the
electrons, and the introduction of an 
homogeneous background (the ''pudding''), to fulfill electro-neutrality.
This picture, although obsolete in the atomic realm, has nevertheless 
attracted the interest of mathematicians, physicists
and biologists alike, due to its 
relevance in particular for ionic ordering
at interfaces \cite{Leiderer}, for the behaviour of colloids self-assembled at 
the edge of emulsion droplets (colloidosomes, see e.g. \cite{Weitz}), for the
study of one-component plasmas and their experimental realizations (electrons 
on a liquid Helium surface \cite{KonoSan}) or
for understanding viral morphology \cite{CK}. 
The Thomson problem reappeared recently in sheep's clothing
in different contexts, from the screening effects in hydrophobic
polyelectrolytes \cite{C09}, to the behaviour of Coulomb balls
(identical particles confined in an harmonic trap \cite{Wrighton}),
including hydrogels \cite{Holm}, where the uptake of counter-ions
by a cross-linked polymer network (the ``pudding'') is the key
feature leading to the expansion of the network by osmotic
pressure, hence the capability to absorb large quantities of water. 
At variance with Thomson's preoccupation where ground state configurations
were under scrutiny,
those articles were concerned with the finite temperature behaviour
($T\neq 0$).
At high to moderate temperatures, mean-field theory provides
a trustworthy framework, and allows to obtain some analytical results
\cite{C09}. The regime of low temperatures (large couplings,
a notion to be specified below) is
more elusive, and it will be our primary objective in the present contribution.
Several analytical predictions will be derived, including 
scaling laws for two important quantities characterizing
the screening properties. These predictions will be tested against
numerical simulations, that also give access to detailed microscopic
information concerning the structure.


We start by defining the model and introducing a relevant coupling parameter.
We consider a single permeable and spherical globule of radius $R_g$ and charge 
$-z q Z_g$ (referred to as the background), 
surrounded by its $Z_g$ counter-ions of charge $z q$, where
$q$ is the elementary charge and $z$ the ionic valency. It should be 
stressed that the counter-ions can penetrate but also {\em leave} the
homogeneously charged globule, which is an important difference
with Thomson's original formulation. Without loss of generality,
the globule is assumed negatively charged (positive counter-ions).
In such a salt-free system, a confining cell 
is required  to avoid evaporation of all counter-ions; its radius is 
denoted $R_c$. 
A key quantity is the ``globule total uptake''  
charge $-\Zeff$, which includes the background and the 
counter-ions present within the globule \cite{rque95}. 
Hence, $Z_g-\Zeff$ can be viewed as
the number of counter-ions outside the globule. 
Following~\cite{Levin,Wrighton}, we define the coupling (plasma) parameter 
$\Gamma $ as the ratio between the characteristic electrostatic energy of a 
counterion-counterion interaction, $z^2 q^2 / (4 \pi \epsilon \delta)$,
and the thermal energy, $kT$. Here $\epsilon$ is the solvent 
dielectric constant, and
the distance $\delta$ is taken as the ion sphere radius:  $ Z_g \delta^3 = R_g^3$. 
Introducing the Bjerrum length $\ell_B=q^2/(4 \pi \epsilon kT)$, 
which is about 7\,\AA\ 
in water at room temperature, we get 
\be
\Gamma \,= \, \frac{z^2 q^2}{4 \pi \epsilon \delta \,kT} \,=\, 
Z_g^{1/3}\,\frac{z^2 \ell_B}{R_g}.
\label{eq:Gamma}
\ee
For $\Gamma<1$, mean-field Poisson-Boltzmann theory \cite{Levin} 
provides an accurate
description, while the strong coupling regime corresponds to $\Gamma>1$. 
It has been shown in \cite{C09} that within the non-linear mean-field
regime \cite{rque2}, counter-ion penetration into the globule lead to 
$\Zeff \propto \sqrt{Z_g}$
for large enough bare charge $Z_g$. For the purpose of comparison with 
numerical calculations in the strongly coupled regime, this quantity
needs suitable rescaling, and we define
\be
\Pi \,=\, \Zeff\, \frac{z^4 \ell_B^2}{R_g^2},
\ee
which only depends on $\Gamma$ \cite{rque20}. Within mean-field,
we then have  $\Pi\propto \Gamma^{3/2}$ up to a prefactor of order 1,
and in addition, the relation $n \propto \exp(-z e \phi/kT)$ between
counter-ions density $n(\mathbf{r})$ and the 
local mean electrostatic potential $\phi(\mathbf{r})$ \cite{Levin},
allows us to relate the reduced charge $\Pi$ to the characteristic
decay length of $n(r)$ at $r=R_g$.
To this end, we define the positive quantity
\be
S \,=\, -z^2 \ell_B \, \frac{d}{dr}\biggl|_{r=R_g} \log n(r),
\label{eq:defS}
\ee
that can be viewed as the reduced (inverse) decay length
of the counter-ion profile in the globule vicinity.
Gauss theorem implies $S=\Pi$, again within irrelevant prefactors:
the reduced decay length $S^{-1}$ is inversely proportional to 
the reduced total charge, that is itself an increasing function 
of the background charge $Z_g$.

How are the previous results affected in the strong-coupling regime?
To make analytical progress when $\Gamma \gg 1$, we take 
advantage of the caging of particles that takes place 
under their strong mutual repulsion \cite{Hasse,rque100}. This basic feature of
strongly-coupled Coulomb or Yukawa plasmas is at the root of the
quasi-localized charge approximation, that has proved useful
for the determination of dynamic quantities \cite{QLCA}: the charges
inside the globule are trapped around local potential minima, and therefore
oscillate at a frequency close to the Einstein value 
$\omega_E^2 =  n_0 z^2 q^2/(3 m \epsilon) $ \cite{Einstein,rque}, 
where $n_0 = Z_g/(4 \pi R_g^3/3)$ is the background density, 
and $m$ is the counter-ion mass. The potential felt locally by a counter-ion 
then reads
\be
U(x) \,=\, \frac{1}{2} \,\frac{z^2 q^2 n_0}{3\,\epsilon} \, x^2 = 
\frac{1}{6} \,\frac{kT \,x^2}{\lambda^2}
\ee
where $x$ stands for the deviation from potential minimum,
and $\lambda = [z^2 q^2 n_0/(\epsilon kT)]^{-1/2}$ can be thought of as
a Debye length. The typical cage size is given by $\delta$,
as required from local electro-neutrality, and likewise,
those cages located near the boundary of the globule ($r=R_g$)
are centered at $r=R_g-\dmax$. 
We expect $\dmax$ and $\delta$ 
to scale accordingly, and more precisely, we write $\dmax=\alpha \delta$,
where $\alpha$ will be an important quantity for what follows.
Due to the repulsion of neighboring counter-ions, we anticipate 
$\alpha<1$ (the outer layer of confined ions is ``pushed''
towards the boundary $r=R_g$, by a mechanism reminiscent
of depletion in hard core systems). This is precisely the scenario 
at work in the ground state ({\it i.e.}\/ at infinite $\Gamma$), where several
approximate expressions have been proposed for $\alpha$ \cite{Hasse}.
Consistent with these approximations and with numerical simulations
that report $0.73<\alpha<0.77$ \cite{Hasse},
we will take $\alpha=3/4$ for our large coupling expansions.

We are now in position to compute the 
uptake charge of the globule, from  the number of counter-ions 
that are able to escape their cage. At large $\Gamma$, only 
those cages located near the globule boundary can loose particles; 
there are $(R_g/\delta)^2$ such cages, so that 
\begin{eqnarray}
\Zeff &\simeq& \frac{R_g^2}{\delta^2} \, \int_{\dmax}^\infty \, 
\frac{dx}{\lambda} \, \exp\left(- \frac{x^2}{6 \lambda^2}\right) \\
&\simeq&  \frac{R_g^2}{\delta^2}\, \frac{\lambda}{\dmax} 
\exp\left(- \frac{\dmax^2}{6 \lambda^2}\right).
\end{eqnarray}
Since $\delta^2/\lambda^2  = 3 z^2\ell_B/\delta = 3 \Gamma$, 
and going from $\Zeff$ to its rescaled form $\Pi$, we obtain 
\be
\Pi \, \propto \, \Gamma^{3/2} \exp(-\alpha^2 \Gamma /2).
\label{eq:Pi_scaling}
\ee
This shows, under strong coupling and at variance
with mean-field, that the uptake charge actually
{\em decreases} upon increasing the globule charge; 
furthermore, it is noteworthy that our argument,
valid at large $\Gamma$, also reproduces the 
mean-field small $\Gamma$
behaviour with a power law of exponent 3/2.

We now seek a more microscopic information and attempt to 
predict the radial dependence of the counter-ion profile $n(r)$ 
outside the globule. When a counter-ion approaches the globule,
it polarizes the trapped ions in their cages, and in turn feels
the potential $V(r)$ thereby created (the more obvious $\Zeff/r$ contribution 
appears to be sub-dominant, see below). More specifically,
the test particle located at $\mathbf{r}$ creates a field 
$\mathbf{E}=-zq (\mathbf{r}-\mathbf{x})/(4 \pi \epsilon 
|\mathbf{r}-\mathbf{x}|^3)$
at point $\mathbf x$ where a counter-ion located inside the globule
will be displaced from its equilibrium position, creating a dipole 
moment $\mathbf{p} = 3 z^2 q^2 \lambda^2 \,\mathbf{E}/kT$ \cite{rque4}.
The contribution of this dipole to the potential $V$ felt by the test
charge is
$\mathbf{p}\cdot (\mathbf{r}-\mathbf{x})/(4 \pi \epsilon 
|\mathbf{r}-\mathbf{x}|^3)$,
an expression that we have to integrate over all cages of counter-ions
(one dipole for each cage); moreover, 
for consistency with our previous argument
with outer cages centered at a radial position 
$R_g-\dmax$, and gathering expressions, we have
\be
V(\mathbf{r}) \simeq - n_0 \frac{3 z^3 q^3 \lambda^2}{(4\pi\epsilon)^2 kT} \,
\int_{|\mathbf{x}|\leq R_g-\dmax} \frac{1}{|\mathbf{r}-\mathbf{x}|^4} \, d^3 \mathbf{x}.
\ee
To obtain the dominant contribution, in the vicinity of $R_g$, 
we neglect the curvature of the globule, which gives
\be
\frac{-zq V(r)}{kT}\,\propto \, \frac{\ell_B}{r-R_g+\dmax} 
\quad \hbox{with} \quad r=|\mathbf{r}|.
\ee
The corresponding density profile 
follows from $n(r) \propto e^{-V/kT}$, which yields the dominant behaviour
$\log n \propto \ell_B/(r-R_g+\dmax)$. This allows not only to compute the scaling
parameter $S$ but also to propose a scaling function for $n(r)$. 
Indeed we obtain here $S\propto z^2 \ell_B^2 / \delta^2$
from the definition (\ref{eq:defS}), i.e. 
$S\propto \Gamma^2$, and 
\be
\frac{1}{\Gamma} \log\frac{n(r)}{n(R_g)} \,\propto\, {\cal F}(\zeta) 
\,=\,  [(\zeta+\alpha)^{-1} - \alpha^{-1} ]
\label{eq:scaling_n} 
\ee
with $\zeta = (r-R_g)/\delta$.
As a consequence, if the scaling relation (\ref{eq:scaling_n})
holds, an important test for the consistency of our approach is to
recover the same value of $\alpha$ as in expression 
(\ref{eq:Pi_scaling}), close to 3/4.
To summarize our analytical findings, we obtained,
in addition to the explicit scaling form (\ref{eq:scaling_n}), that while 
$\Pi \propto S \propto \Gamma^{3/2}$ in the non-linear mean-field
regime (meaning the limit of large bare charges $Z_g$ within
Poisson-Boltzmann theory), strong coupling leads to $S \propto \Gamma^2$
and an uptake charge (background plus counter-ions present within the 
globule) $\Pi \, \propto \, \Gamma^{3/2} \exp(-\alpha^2 \Gamma /2)$.
The slope ($S$) increases with $\Gamma$ steeper than in mean-field, and 
ionic correlation effects lead at large $\Gamma$ to a decrease of 
the charge $\Pi$.

\begin{figure}[ht]
\begin{center}
\includegraphics[clip=true,width=8cm]{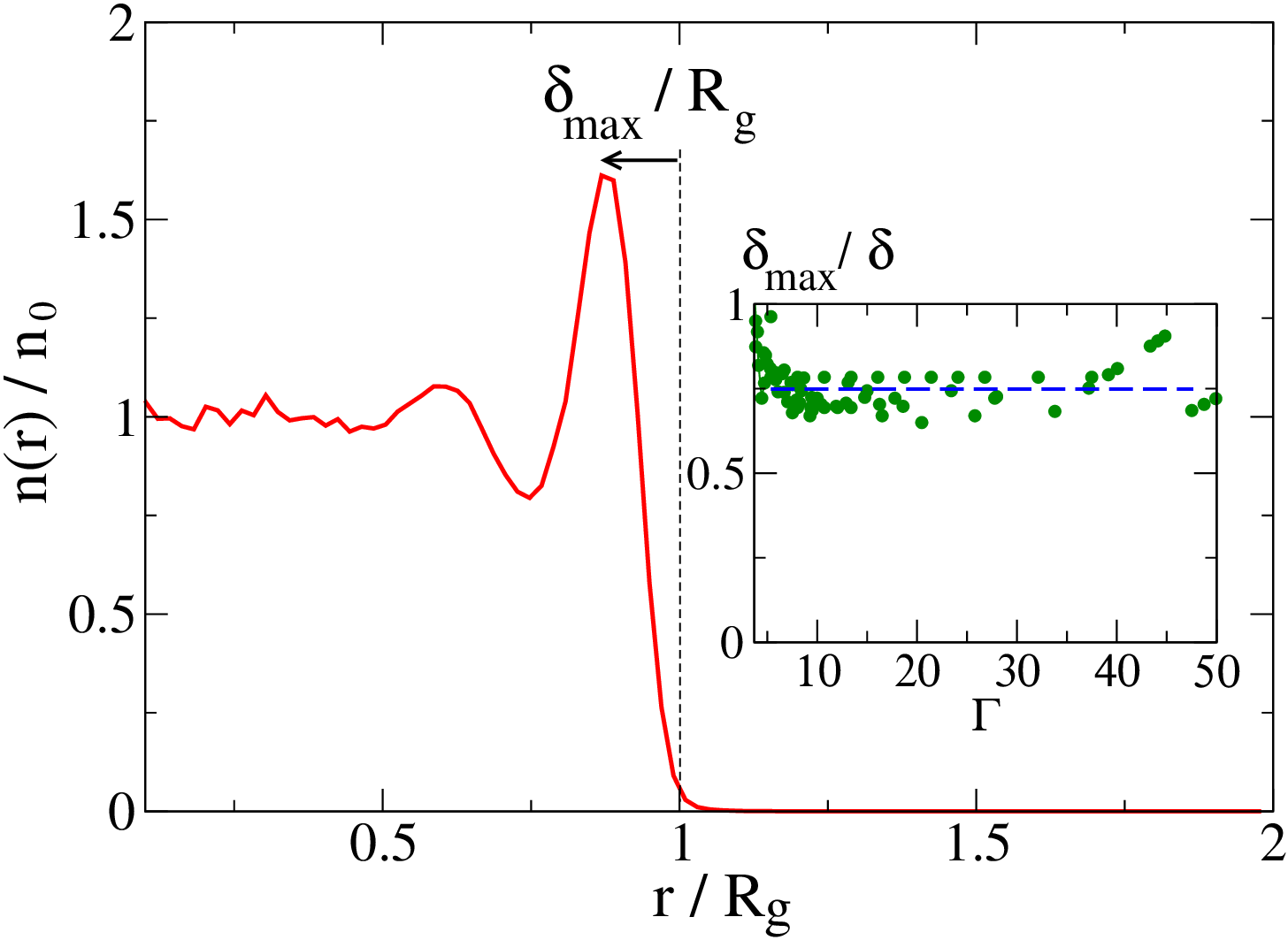}
\vglue -0.25cm
\caption{Counter-ion density profile $n$ normalized by background density $n_0$
as a function of distance from the globule center (normalized by 
the globule radius $R_g$). Here $\Gamma= 13$,
$R_c/R_g=2$ and $Z_g=200$. The density peak location defines
the distance $\dmax$, that is shown in the inset at various couplings,
 different cell radii ($R_c/R_g=2, 4$ and 8 and $Z_g$ from 50 to 3000. 
The dashed line shows the value 3/4, used for $\alpha$ throughout this work.
}
\label{fig:density_alpha}
\end{center}
\end{figure}

\begin{figure}[ht]
\begin{center}
\includegraphics[clip=true,width=8cm]{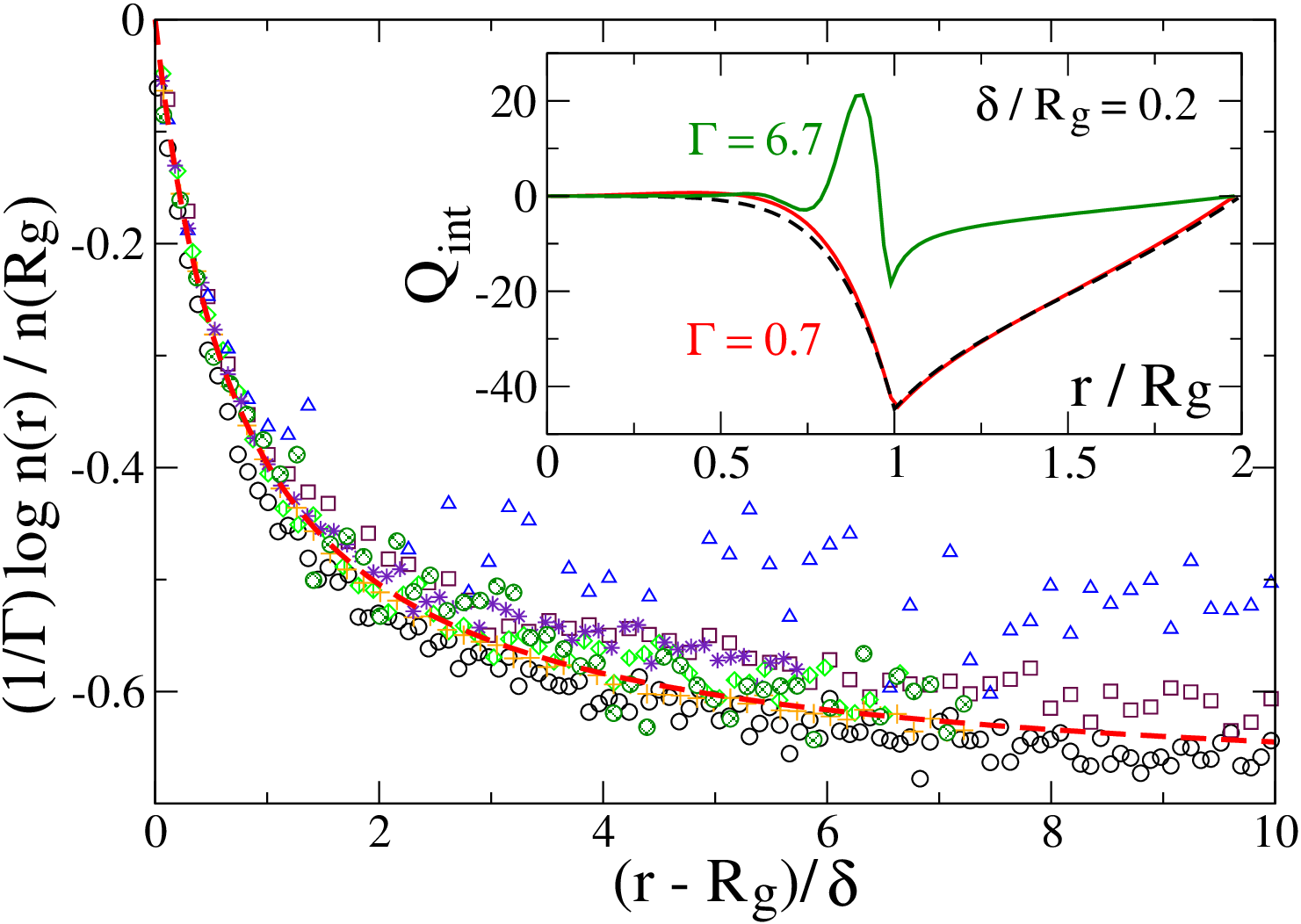}
\end{center}
\caption{
Plots of rescaled counter-ion profiles in the vicinity of the
globule, $\Gamma^{-1} \log[n(r)/n(R_g)]$, as a function of 
$\zeta=(r-R_g)/\delta$.  The different symbols correspond 
to different values of $\Gamma$, from 8 to 24. The thick dashed curve
is for the function ${\cal F}(\zeta)/2$ where $\cal F$ is
defined in Eq. (\ref{eq:scaling_n}), 
with a value $\alpha=3/4$. 
Inset:  Integrated charge $\Qint$ vs radial distance, at weak and strong 
couplings.
For $\Gamma<1$, $\Qint$ is always of the same sign as the background,
while over-charging is observed at strong couplings. The dashed
line shows the mean-field result.
}
\label{fig:n_scaling}
\end{figure}

\begin{figure}[htb]
\begin{center}
\includegraphics[clip=true,width=8cm]{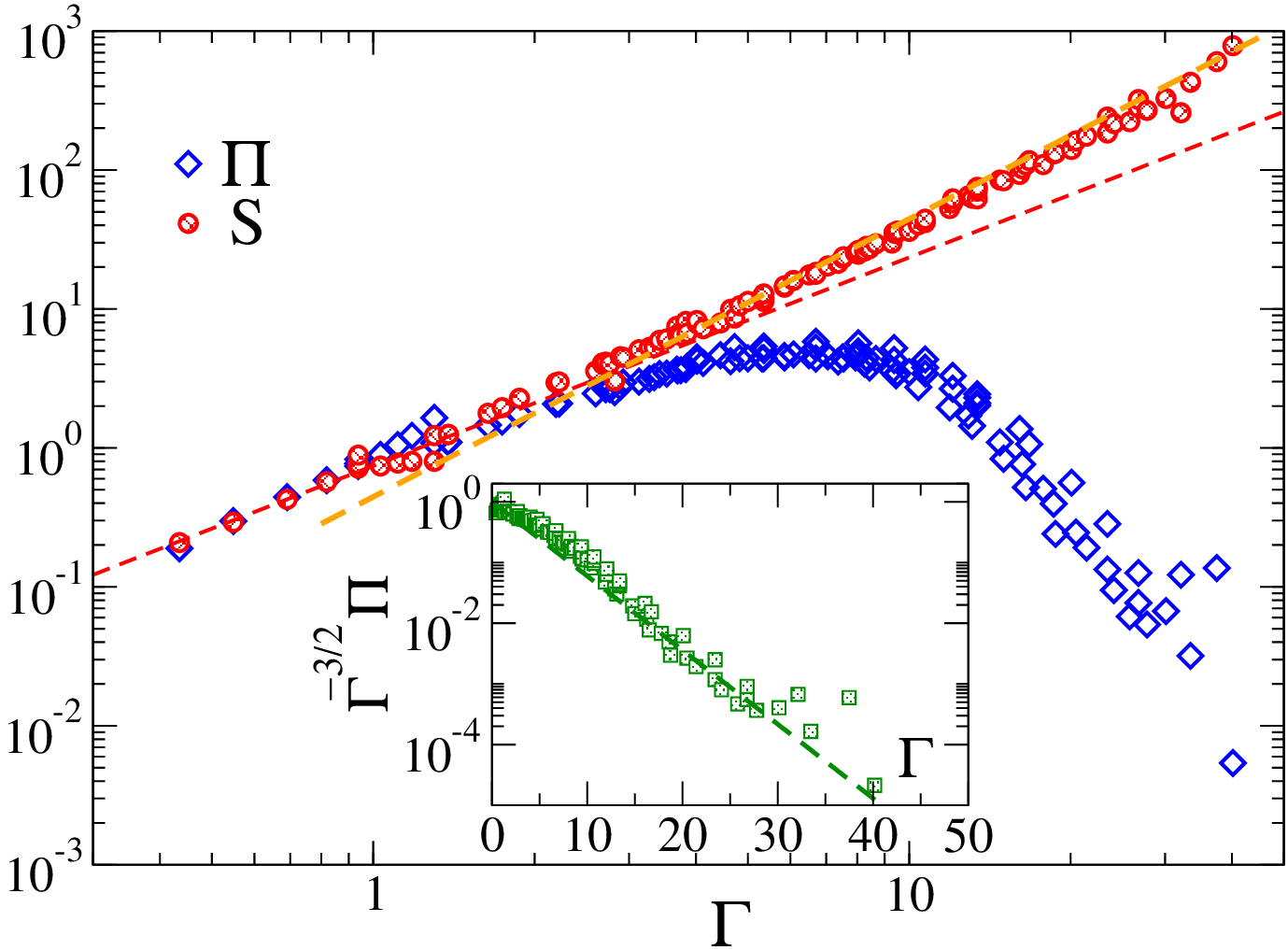}
\end{center}
\caption{Rescaled counter-ion density slope $S$ defined in 
Eq. (\ref{eq:defS}), and
reduced uptake charge $\Pi = \Zeff\, z^4 \ell_B^2/R_g^2$,
as a function of coupling parameter $\Gamma$. 
The dashed lines have slope 3/2 (to evidence the mean-field behaviour)
and 2. Here $S$ has been divided by 2 to have $S\simeq \Pi$ at low 
$\Gamma$.
The inset shows $\Pi/\Gamma^{3/2}$ as a function 
of coupling, on a linear-log scale. The line has slope $-\alpha^2/2 \simeq -0.28$.}
\label{fig:S_Pi}
\end{figure}

To put these predictions to the test, we have performed Monte Carlo
simulations, where the counter-ion interact through the exact 
Coulomb law and feel the background charge of the globule,
the whole system being furthermore enclosed in a larger sphere of radius
$R_c$ \cite{rque20}. 
Following the analytical treatment, we located the position of
counter-ionic density peak close to the globule edge, which
defines $\dmax$, see Fig. \ref{fig:density_alpha}. 
The inset of this figure shows that
the ratio $\alpha=\dmax/\delta$ does not depend on the coupling strength,
and remains close to its ground state limit $\alpha\simeq 3/4$. 
Recovering the proper value of $\alpha$ at large $\Gamma$ can be
viewed as an assessment of the validity of our simulations,
and we can then proceed with the explicit check of the
scaling form Eq. (\ref{eq:scaling_n}). It can
be seen in Fig. \ref{fig:n_scaling} 
that the different curves exhibit good collapse at different $\Gamma$, 
and that the function ${\cal F}(\zeta)$
captures the density decay in the external vicinity of the
globule ($\zeta >0$). While our argument above only provides the
relation $\Gamma^{-1} \log[n(r)/n(R_g)] \propto {\cal F}(\zeta)$,
a more refined analysis indicates that the prefactor is close
to 1/2 \cite{future}, so that we have plotted ${\cal F}(\zeta)/2$
in Fig. \ref{fig:n_scaling}.
We have also computed the values of $S$ and $\Pi$ in the simulations. 
They are shown in Fig.  \ref{fig:S_Pi}, which fully corroborates 
the analytical scaling behaviours. 
First, at small $\Gamma$, we have the mean-field behaviour 
$S \propto \Pi \propto \Gamma^{3/2}$, while at larger couplings,
$S\propto \Gamma^2$ and $\Pi$ becomes non monotonous. 
The detailed behaviour of $\Pi$ versus $\Gamma$ provides  
a stringent test for our arguments: as shown in the inset of
Fig. \ref{fig:S_Pi}, the dependence of $\log(\Pi/\Gamma^{3/2})$
on $\Gamma$ is linear at large $\Gamma$, with a (negative) slope compatible
with the predicted value of $\alpha^2/2=9/32$. We see that a 
unique value of $\alpha$, inherited from ground state properties,
accounts for the behaviour of the density profile together with more
global quantities like the uptake charge.

The previous considerations provide a detailed description 
for the ionic density profile outside the globule. The behaviour 
for $r<R_g$ is more intricate, and has been analyzed in Monte
Carlo. A quantity of interest is the total integrated charge $\Qint(r)$
inside a sphere of radius $r$. By definition, $\Qint(R_g)=-\Zeff$ and $\Qint(R_c)=0$
due to electro-neutrality. Within mean-field, it is interesting to 
note that $\Qint$ is always of the same sign as the background
(here, negative). One can also note that the salt-free nature of our system 
imposes $\Qint(R_g)<0$, and more precisely, that $\Qint(r)<0$ for all
$r \geq R_g$. Hence, a true overcharging cannot be observed in our
salt-free system \cite{Shklovskii, Levin}.
However, for $\Gamma>1$, the inset of 
Fig. \ref{fig:n_scaling} shows that $\Qint(r)$
exhibits a range of distances where it is positive: 
we refer to such a possibility as a {\em local} over-charging,
and we note that it 
is somehow reminiscent of its counterpart occurring for impermeable
colloids \cite{Shklovskii, Levin}. Indeed, as for impermeable colloids,
it is possible to prove that local overcharging is precluded within mean-field
theories \cite{T00,future}, so that when present, 
it is a manifestation of ionic
correlations, that become prevalent at $\Gamma>1$. 
The electrophoretic consequences of this over-charging effect
are unclear, and left for future study.
Incidentally, the close agreement between Monte Carlo and mean-field results 
at small $\Gamma$ (see inset of Fig. \ref{fig:n_scaling}) can be seen 
as assessing the validity of the numerical
methods employed.

To summarize, we investigated the screening properties of a 
uniformly charged spherical globule, neutralized by point
counter-ions. Invoking a quasi-localized charge argument
at strong Coulombic coupling $\Gamma$, we obtained analytically the 
counter-ion density profile $n(r)$ outside the globule, together with
two more global quantities: one, denoted $S$, follows from $n(r)$ 
and is its characteristic inverse decay length in the globule edge vicinity
($r=R_g$); the second quantity, $\Pi$,
stands for the reduced total charge inside the globule and quantifies
the counter-ion uptake. At small couplings, $S$ and $\Pi$ 
coincide and scale as $\Gamma^{3/2}$. Strong ionic correlations,
on the other hand, were shown to lead to a departure of both
quantities: the slope $S$ becomes steeper (algebraic increase in $\Gamma^2$)
and the counter-ion uptake
is much more efficient, leading to a charge 
$\Pi$ that
{\em decreases} upon increasing $\Gamma$ (which can be obtained increasing 
the globule bare charge $Z_g$): the algebraic mean-field increase
turns into an exponential decrease, see Eq. (\ref{eq:Pi_scaling}).
Our scaling predictions are free of adjustable parameters,
and make use of known ground state properties \cite{Hasse} 
(that fix the parameter $\alpha=\dmax/\delta$).
These predictions were corroborated by Monte Carlo simulations, that provide
the exact static properties of our system at arbitrary $\Gamma$,
and that furthermore revealed an over-charging effect
that is absent within mean-field scenario.
We have treated the phenomenon of caging at a rather simple
level, that turns out to be sufficient to capture the interesting 
violations of mean-field, driven by ionic correlations.
In addition, while the mean-field theory applies schematically
for $\Gamma<1$ and the strong-coupling arguments cover the 
range $\Gamma>10$, a quantitative understanding of the cross-over
region at moderate couplings presumably requires intermediate 
approaches in the spirit of Refs. \cite{Burak}.

Our work opens interesting venues for future studies.
First, our predictions can be tested experimentally,
in the spirit of the experiments reported in \cite{Essafi}.
For the corresponding hydrophobic polyelectrolytes,
we estimate that $\Gamma \simeq z^2$ {\em at room temperature},
where $z$ is the valency of the counter-ions. 
Consequently, with trivalent ions, one has $\Gamma \simeq 9$,
which is the regime where mean-field no longer applies,
and our strong-coupling predictions take over, see Fig. \ref{fig:S_Pi}.  
Second, the phase behaviour of an ensemble 
of such globules is unknown, together with the effects
of an added electrolyte. Finally, the response to external 
perturbations, both static (sedimentation) or dynamic (electrophoresis),
should provide a relevant ground to investigate the
signature of strong Coulombic correlations.



\begin{thebibliography}{99}
\bibitem{rque2}
Note that the weak-coupling limit $\Gamma<1$
is compatible with strong non-linear effects, which occur 
within mean-field whenever 
$Z_g \ell_B/R_g \gg 1$ (which is easily met), 
provided $\ell_B\ll R_g$, see Eq. (\ref{eq:Gamma}).
\bibitem{rque95}
We avoid the terminology ``effective'' for $\Zeff$, which has a different meaning
(charge extracted from the far-field behaviour).
\bibitem{rque20}
At large $\Gamma$, the particular choice of $R_c$ does not affect 
our predictions --as can be seen in Fig. \ref{fig:density_alpha}--
as long as it is finite.
\bibitem{rque4}
A simple argument provides the needed 
polarizability of a cage: for an ion displaced by a quantity 
$\delta\mathbf{r}$
from its equilibrium position by an external field $\mathbf{E}$, 
we have the force 
balance condition 
$m \omega_E^2 \delta\mathbf{r} = z q \mathbf{E}$, which yields
the dipole moment $\mathbf{p}= z q \,\delta\mathbf{r} = z^2 q^2 \lambda^2
\mathbf{E}/(kT)$.
\end{thebibliography}
\end{document}